# Atomistic Modelling of Thermal-Cycling Rejuvenation in Metallic Glasses

Baoshuang Shang[1,2], Weihua Wang[3,2], Alan Lindsay Greer[4], and Pengfei Guan[1,*]

[1]*Beijing Computational Science Research Center, Beijing 100193, China*

[2]*Songshan Lake Materials Laboratory, Dongguan 523808, China*

[3]*Institute of Physics, Chinese Academy of Sciences, Beijing 100190, China*

[4]*Department of Materials Science and Metallurgy, University of Cambridge,*

*27 Charles Babbage Road, Cambridge CB3 0FS, UK.*

Cycling of a metallic glass between ambient and cryogenic temperatures can induce higher-energy states characteristic of glass formation on faster cooling. This rejuvenation, unexpected because it occurs at small macroscopic strains and well below the temperatures of thermally induced structural change, is important, for example, in improving plasticity. Molecular-dynamics simulations elucidate the mechanisms by which thermal cycling can induce relaxation (reaching lower energy) as well as rejuvenation. Thermal cycling, over tens of cycles, drives local atomic rearrangements progressively erasing the initial glass structure. This arises mainly from the heating stage in each thermal cycle, linked to the intrinsic structural heterogeneity in metallic glasses. Although, in particular, the timescales in MD simulations are shorter than in physical experiments, the present simulations reproduce many physically observed effects, suggesting that they may be useful in optimizing thermal cycling for tuning the properties of metallic glasses and glasses in general.

*pguan@csrc.ac.cn

For any type of glass of given composition, its properties can vary widely, depending on the structure reached by varying the production process and subsequent treatments[1]. For metallic glasses (MGs), annealing-induced relaxation (ageing) can lead to severe embrittlement, and there is therefore a focus on the reverse process of rejuvenation. Rejuvenation can be induced by mechanical deformation[2–6], by reheating and rapid quenching[7], or irradiation[8], but among such methods, cryogenic thermal cycling (CTC)[9] is particularly straightforward to apply. Since the original report[9], there have been many experimental studies of CTC applied to MGs. These have shown effects on the enthalpy[10–25], on density[11–17, 21, 26], on mechanical properties[10–22, 24–31] and on structure[19, 20, 24–27, 30, 32–34]. The most dramatic beneficial effects of rejuvenation are on the fracture behaviour: fracture toughness and impact toughness roughly doubled[18, 28], and toughness (critical strain-energy release rate) increased by more than a factor of five[18]. The underlying atomistic mechanisms of the effects of CTC clearly merit investigation.

In MGs, the atomic nearest-neighbour coordination shells are diverse, and it is increasingly recognized that there is significant structural and dynamic heterogeneity over a range of length-scales. Accordingly, the strains associated with thermal expansion/contraction must have a significant non-affine component[35, 36] that must be associated with the development of internal stresses. While CTC was initially aimed at rejuvenation of MGs, other work[27, 37] has shown that this treatment can also induce relaxation and oscillatory (rejuvenation-relaxation) behaviour. Relatively easy rearrangement, associated with non-directional metallic bonding, may make the effects of thermal cycling (TC) particularly significant in MGs, but the non-affine nature of thermal strains applies across all classes of non-crystalline material, and is relevant (though scarcely studied so far) for the interpretation of the evolution of their structure and properties whenever temperature changes are involved.

For MGs, molecular dynamics (MD) simulations have been useful in understanding the links between structure and dynamics, and specifically the role of structural heterogeneity[38]. MD simulations have been used to study the effects of TC on a model Lennard-Jones (L-J) glass[39–41]; in this series of studies, the temperature variation with time had a triangular waveform, i.e. there was no hold-time at the upper and lower temperatures. The TC had an amplitude as high as 2–92% of the glass-transition temperature $T_g$. Predominantly, the effect of TC was to lower the potential energy of the glass, and the extent of this relaxation was maximum for intermediate values of the TC amplitude. In the glass formed at the lowest cooling rate and subjected to the maximum amplitude of TC, it was just possible to detect rejuvenation. The MD simulations showed that the extent of the stress overshoot on a tensile stress-strain curve was clearly greater for lower energy states. The simulations also allowed mapping of the distribution of clusters within which the non-affine displacements induced by TC are concentrated.

In the present work, we simulate TC of an actual glass, $Cu_{50}Zr_{50}$ (composition in at.%), and apply heating and cooling, at $10^{14}$ K s$^{-1}$, that is essentially instantaneous. We find clear rejuvenation as well as relaxation, and we map the boundary between these effects. MD simulations (generally, and in the present work) involve timescales much shorter than those typical in physical experiments. Nevertheless, we show that MD can assist in understanding the link between TC conditions and property changes. In particular, the simulations can identify the relevant length-scale of heterogeneity (a question not addressable in the original work on CTC[9]), and can identify the stage in TC mainly responsible for rejuvenation.

## Results

**Potential energy variation during thermal cycling.** $Cu_{50}Zr_{50}$ (at.%) metallic glasses are simulated using classical MD (see Methods). Glasses formed by quenching from the liquid state at different rates $Q$ are subjected to TC between $T_a$ (400 K) and $T_b$ (mostly 1 K) with hold times of $t_a$ and $t_b$ at these temperatures. The potential energy $E$ is monitored, and Fig. 1a shows a case where its value (initially $E_0$ at the end of the quench) is higher after each cycle. The effect of the overall processing (*N* cycles) is characterized by the value at the end of the holding time $t_a$ at the final temperature (400 K).

A higher $Q$ gives an initial glass with a higher $E_0$ (Fig. 1b). Values of $E$, after up to 60 cycles, are shown for $t_a = 10$ ps and $t_a = 200$ ps; for a given $t_a$, independent of $E_0$, all the potential energies converge to a common value after about 40 cycles (Fig. 1b). Following the general interpretation of the effects of energetic processing[1], this value $E_s$ is taken to represent a steady state in which structural damage and repair rates are in balance. Comparing $t_a = 10$ ps and 200 ps, the longer time allows for more relaxation and therefore gives a lower $E_s$. For 10 ps, all the glasses show rejuvenation on cycling; for 200 ps, the glass formed at $10^{12}$ K s$^{-1}$ stays at roughly the same energy, and the most rapidly quenched glass ($Q_5 = 10^{13}$ K s$^{-1}$) undergoes relaxation.

For a given treatment, a more relaxed glass shows slower rejuvenation (Fig. 1b, and Supplementary Fig. 1a where the energy changes are compared from a common starting point). This is consistent with the suggestion[35] that the origin of the rejuvenation is the heterogeneity of the glass structure; a more relaxed initial glass (formed on slower cooling) would have a more uniform structure. A more direct test of the effect of uniformity is obtained by MD simulations of B2 CuZr crystal. In this case, crystallographic symmetry dictates that the thermal strains are affine. The potential energy indeed shows no changes as a result of TC (see Fig. 6c and Fig. 7b later).

For a given initial glass, the effect of $t_a$ on the $E_s$ induced by TC is shown further in Supplementary Fig. 1b: $E_s$ depends strongly on $t_a$ and intersects the $E_0$ values of the as-quenched glasses (Fig. 2). A given as-quenched state can thus be rejuvenated or relaxed by TC (Fig. 2), just as seen in experiment[27]. Other TC parameters being equal, there is a critical value $t_c$ (arrowed): for $t_a < t_c$ there is rejuvenation, and for $t_a > t_c$ relaxation. The ultimate energy change on TC ($\Delta E_s = E_s - E_0$) can be mapped as a function of $t_a$ and of the initial $E_0$ value of a range of glassy samples (Supplementary Fig. 2). Extrapolating the trends on the map, for a typical bulk metallic glass with cooling rate $10^4$ K s$^{-1}$, the holding time at the upper temperature should be shorter than 100 s to achieve rejuvenation.

Upon TC, the model L-J glass cited earlier[39–41] predominantly showed relaxation. The higher of the quench rates used to form the initial glasses in the present work (>$10^{11}$ K s$^{-1}$) overlap the quench rates applied to the L-J system, yet our glasses formed at those higher rates still predominantly show rejuvenation upon TC (Fig. 1b). This contrasting behaviour may reflect the heating and cooling rates ($Q_a$ and $Q_b$) used in our TC; these are some two orders of magnitude higher than the TC rates applied to the L-J glass, and the consequences are explored further below. In some of the TC simulations for the L-J glass, the upper temperature was as high as 92% of $T_g$, and this would also favour overall relaxation.

**Structural and mechanical properties of cycled samples.** Volume-average properties are compared for a glass with a given $Q$ in its as-quenched state and after TC with two different $t_a$ values to achieve steady states that represent relaxation (ageing), and rejuvenation with respect to the as-quenched glass (these three states are indicated on Supplementary Fig. 2). In each sample, the shear and bulk elastic moduli show a distribution of values. As expected, the distributions are shifted upwards in the relaxed sample and downwards in the rejuvenated sample (Fig. 3a). Atomic pair distribution functions show minor differences between the three samples. The changes induced by rejuvenation are opposite to those induced by relaxation (Fig. 3b). The different configurations of nearest-neighbour clusters can be described as Voronoi polyhedra. For the glasses in the present work, the relative populations of the most favoured polyhedra are shown in Fig. 3c. These populations all increase on relaxation and decrease on rejuvenation. In all these respects, the relaxation and rejuvenation achieved by TC are qualitatively the same as those achieved by annealing or by forming the glass at different quench rates.

Of most practical interest is the form of the stress-strain curve for shear deformation (Fig. 3d); in a rejuvenated sample, the stress overshoot can be completely eliminated. The elimination of the overshoot is not seen for the simulated L-J glass[39–41], even when formed by quenching much faster than the maximum rate in the present work. By markedly reducing, or even eliminating, the stress overshoot at yielding, rejuvenation offers the prospect of avoiding strain localization and consequent catastrophic mechanical failure, as has been explored experimentally for rejuvenation achieved by means other than TC[8, 9].

**Vibrational behaviour during TC.** MD simulations are of particular interest to characterize local structures and properties in MGs; studies show that there are low-frequency (soft) modes associated with a distribution of soft spots in the glass[42, 43]. As described in Methods, we focus on the vibrational (phonon) density of states (VDOS) $g(E)$ and the atomic participation ratio $P_i$ in low-frequency (soft) modes (Supplementary Fig. 3).

A glass subjected to TC shows an increase in $\Delta E = E_N - E_0$, accelerating as the heterogeneity increases, before levelling off at steady state (inset, Fig. 4a). The VDOS (Fig. 4a) shows the boson peak characteristic of glasses. This peak shifts to lower energy and is more intense as the number of cycles $N$ increases. These effects are similar to those found for heavy plastic deformation of an MG in an experimental and MD simulation study[44], in which they were associated with rejuvenated regions at shear bands.

The atomic participation ratio $P_i$ is a suitable basis for considering temporal and spatial correlations in the MG. The Pearson correlation coefficient $p_{0,N}$ of $P_i$, between the initial state and after $N$ cycles at any given location shows a clear decay towards zero as $N$ increases (Fig. 4b). The memory of the initial state is progressively lost as the sample reaches the steady state, implying that TC can induce comprehensive structural change (α relaxation) even though the temperature is always far below $T_g$. As the population density of soft spots (and the boson peak intensity) increase with the number of cycles, more rearrangement events are triggered during TC; thus the sample structure changes more and more rapidly. The decrease in correlation of $P_i$ between two adjacent cycles as TC progresses (inset, Fig. 4b) reveals the accelerating increase in the heterogeneity of the MG.

The spatial distributions of $P_i$ for the initial glass ($Q_1 = 10^9$ K s$^{-1}$) and after the first cycle (Figs. 4c, d) show localized regions (in red) where high $P_i$ values indicate soft spots. There are examples of both disappearance and generation of soft spots, but overall there is an increase in their population (Supplementary Fig. 4). With increasing number of cycles, the increased population appears to reach saturation, or perhaps even to decrease slightly.

**The influence of TC parameters.** The parameters describing the temperature profile imposed during TC are shown in Supplementary Fig. 5. Rejuvenation and relaxation are competing processes. Greater overall rejuvenation is therefore achieved if relaxation is restrained. At the upper temperature $T_a$ of TC, structural relaxation of the MG is important in interpreting the results: shortening the holding time $t_a$ at $T_a$ is effective in reducing the extent of relaxation and in realizing a higher $E_s$ (Figs. 1b, 2 and Supplementary Fig. 1b).

Relaxation can also be restrained by lowering $T_b$, shortening $t_b$, and using faster cooling and heating rates. With relaxation reduced by changing these parameters, more net rejuvenation is accumulated in each cycle (Supplementary Fig. 6). Strikingly, even though the structure is already frozen at $T_b$, there is still some local relaxation, indicated particularly by the local kinetic energy or temperature (Fig. 5a). Restricting relaxation during the low-temperature hold, by decreasing $t_b$, the distribution of local temperature $T_{b,loc}$ is more spatially heterogeneous (Figs. 5b–d) at the beginning of the heating process; this heterogeneity could be amplified and could enhance the rejuvenation during heating. The heterogeneity of local temperature $T_{b,loc}$ does not continue to decrease as $t_b$ is prolonged; finally the decrease saturates when the heterogeneity of $T_{b,loc}$ is consistent with the structural heterogeneity. By introducing the quasi-equilibrium process (see Methods), the infinite-$t_b$ state can be obtained. At $t_b = \infty$, the degree of rejuvenation is independent of cooling rate $Q_b$ and lower temperature $T_b$, and correlated only with the heating process.

In a practical implementation of TC, there are likely to be hold times ($t_a$ and $t_b$) at the upper and lower temperatures. We have shown that structural relaxation during these hold times is critical in understanding the overall effect of TC. The effects of such relaxation cannot, of course, be studied if the TC is simulated only with a triangular-wave thermal history as in Refs. [39–41].

**The causes of rejuvenation in TC.** Fixing the conditions (and therefore the likely degree of relaxation) during the upper- and lower-temperature holds, we focus on the effects of the rates of cooling and heating, $Q_b$ and $Q_a$. The potential-energy evolutions in one cycle with slow cooling and heating ($10^{10}$ K s$^{-1}$), and with ultra-fast cooling and heating ($10^{14}$ K s$^{-1}$), are shown in Fig. 6a. For low $Q_b$ and $Q_a$, the potential-energy evolution is almost completely symmetrical on cooling and heating, and the value of $E$ changes only when the temperature is changing. However, for high $Q_b$ and $Q_a$, the symmetry is broken: compared to the $E$ profile at low rate there is little change on cooling, but major changes on heating. At the start of heating at high $Q_a$, there is a lag before $E$ starts to rise rapidly, overshoots its initial value $E_0$, passes through a maximum, and ends the cycle higher than $E_0$. We note that, if the simulation were continued to longer time, $E$ might well decay back to $E_0$; in other words, this rejuvenation may be transient.

The evolution of $E$ at high $Q_a$ implies that the reheating stage plays the key role in TC-induced structural change. It is of interest to examine the correlation between the changes of potential energy $\Delta E_p = E - E_0$ and kinetic energy $\Delta E_k = E_k - E_{k,T_b}$. The increase in potential energy of the glass clearly lags behind the increase in kinetic energy during ultra-fast heating (Fig. 6b, red data points) while there is no similar effect on slow heating (Fig. 6b, blue data points). This suggests that, at least in these MD simulations, the glassy system during ultra-fast heating falls out of "equilibrium" by violating the equipartition theorem (ET); there must be local decoupling of the potential and kinetic energies. In the simulation, the degree of adherence to the ET can be adjusted by choosing different thermostats, and thereby different degrees of rejuvenation are achieved (Supplementary Fig. 7 and the associated discussion in Supplementary Material). By using the Langevin thermostat to force adherence to the ET, rejuvenation of the glass is suppressed (Fig. 6b, orange data points). Using the usual Nosé-Hoover thermostat, and applying TC to the B2 CuZr crystal there is again no rejuvenation (Fig. 6b, green data points). The contrasting effects of TC (rejuvenation or not) are shown in Fig. 6c. Thus, even at high $Q_a$, rejuvenation is achieved only when: (i) there is violation of the ET under the Nosé-Hoover thermostat; <u>and</u> (ii) there is structural heterogeneity, as in the glass but not the crystal. This is direct evidence that an external cause, the violation of the ET during ultra-fast heating, takes on its role through an internal cause, the intrinsic structural heterogeneity, in achieving the rejuvenation in MGs.

This point can be explored further through the atomic rearrangements induced by ultra-fast heating. The local intensity of these rearrangements in a voxel $i$ at temperature $T_i$ is characterized by the activated rate, defined as $R_i \equiv -\Delta F_i / k_B T_i$ (see Methods), where $\Delta F_i$ is the activation energy barrier of the voxel (proportional to the local instantaneous shear modulus[45]), and $k_B$ is the Boltzmann constant. The barrier $\Delta F_i$ is directly determined by the local structural packing, which in a MG is distributed inhomogeneously. The $T_i$ is controlled by the fast out-of-equilibrium heating. Spatial distributions of $R_i$ at the start of the hold at $T_a$ after ultra-fast heating are shown for different cases in Fig. 7. For the B2 CuZr crystal, $R_i$ is controlled by the periodic structure (Fig. 7b, corresponding to the green data points in Fig. 6b), even though there can still be considerable local temperature heterogeneity generated by the NH thermostat. For the glassy state, the $R_i$ is still homogeneous without local temperature heterogeneity, when the Langevin thermostat is used (Fig. 7c, corresponding to the orange data points in Fig. 6b). Thus, consistent with what has already been noted, the intense $R_i$ that leads to rejuvenation upon reheating seems to require both structural and kinetic heterogeneity (Fig. 7d).

The rejuvenated glass (i.e. after TC using the NH thermostat) has the broadest distribution of $R_i$ and the highest frequency of local atomic rearrangements during the hold at $T_a$ (Fig. 8a). We calculated the local non-affinity $\bar{D}_i$ to characterize the evolution of local atomic rearrangements with holding time $t_a$. The relatively strong correlation between the local non-affinity and the local activated rate for all atoms (Fig. 8b) in the rejuvenated case emphasizes that the atomic response is mainly controlled by the local coupling between atomic structure and temperature, which is not found in the crystal state or in the glassy state with the Langevin thermostat. The spatial evolution of the non-affinity (Fig. 9) shows that, even after a long hold at $T_a$, the atomic rearrangements are still correlated with the local activated rate at the start of the upper-temperature hold (circled region in Fig. 7a). Over a long enough time, however, thermal relaxation must erase such rejuvenation effects, and ultimately lead to ageing.

## Discussion

Thermal cycling is known to change the properties of polycrystals of metals that have low crystallographic symmetry and therefore an anisotropic single-crystal thermal expansion coefficient[46, 47]. For such polycrystals, the relevant microstructural length-scale is clearly the grain diameter. It is now accepted that, well below the temperatures at which relaxation annealing would be conducted, thermal cycling, specifically CTC, is a possible method of changing the structure and properties of MGs.

From the earliest study of the effects of CTC on MGs[9, 35], it has been supposed that heterogeneity in the glass must play a role, but, in contrast to the polycrystalline case, the relevant length-scale has not been clear. While there are now many measurements of CTC-induced property changes in MGs, these do not give any indication of the length-scale of the relevant heterogeneity. The MD simulations in the present work do permit heterogeneity to be visualized in the spatial distributions of the local values of: the atomic participation ratio $P_i$ (Fig. 4c,d and Supplementary Fig. 4), the temperature $T_{b,i}$ (Fig. 5b–d), the activated rate $R_i$ (Fig. 7d), and the non-affinity $D_i$ (Fig. 9a–d). As already noted, these various parameters are strongly correlated. Thus, as seen in these figures, they share a common length-scale: this is found to be 0.5–2.0 nm. We suggest that this length-scale is relevant for understanding TC effects in MGs.

In the TC simulations for the L-J glass, the distributions of atoms showing large non-affine displacements were mapped[39, 41]. In early cycles of the L-J glass formed by quenching at the highest rate, large non-affine displacements were dispersed throughout the sample, but in all other cases, the atoms showing such displacements were dispersed as small clusters. The distribution and size of these clusters are similar to, for example, those of the local participation ratio $P_i$ (Fig. 4c,d).

This length-scale is below the resolution limit of common imaging techniques. For electron microscopy, for example, we note that the thickness of the electron-transparent thin foil is 1 to 2 orders of magnitude greater than the heterogeneity size – accordingly, the detection of structural heterogeneity is hindered by projection effects. Nanoprobe studies of MG surfaces do suggest heterogeneity on the scale down to about 1.5 nm [48], broadly consistent with the present work. Length-scales of tens of nm[49], or even tens of μm[18], have also been suggested to describe the heterogeneity in MGs, but such scales cannot be studied with the simulation methods in the present work. If the length-scale of the heterogeneity giving TC effects is 0.5–2.0 nm, that is favourable for the potential application of TC to treat MG components in MEMS and NEMS devices.

An experimental study of a $Cu_{46}Zr_{46}Al_7Gd_1$ BMG[50] found that TC has a strong rejuvenation effect on some properties (notably the initial yield load in nanoindentation, $F_y$) and only a weak effect on others (e.g. hardness, $H$). The relaxation effect of annealing can be reversed by TC for $F_y$, but not for *H*.  These effects could be consistently interpreted by considering the MG to contain soft spots in a relatively rigid matrix:  TC affects mainly the soft spots, increasing both their population density and the ease of shear within them.  It was inferred that a property such as $F_y$ is strongly affected by the distribution of soft spots, while *H* depends mainly on the rigid matrix, which is barely affected by TC.  The soft spots were associated with shear-transformation zones, and therefore directly connected with improved plasticity[50].

The present work is remarkably consistent with this picture.  As noted above, the simulations give spatial distributions of $P_i$, $R_i$ and $D_i$ that all show soft spots.  It is clear that TC does specifically affect the nature and population of those soft spots.  Furthermore, TC is shown (Fig. 3d) to affect the onset of plastic flow much more than its continuation, consistent with the different effects measured[50] for $F_y$ and *H*.

It is evident both from physical measurements[27, 37] and from the simulations in the present work (Fig. 2, Supplementary Fig. 2) that the effect of TC on MGs is delicately balanced between relaxation and rejuvenation.  The dynamic nature of that balance can be better understood by noting that even when, for example, there is a strong overall rejuvenation, soft spots are disappearing as well as appearing (Fig. 4c,d).

In the present work, our aim has been to advance understanding of the effects of TC on the structure and properties of MGs, and in particular how a given MG can be rejuvenated to states of higher energy or aged to lower energy. We have conducted molecular-dynamics (MD) atomistic simulations of cycling $Cu_{50}Zr_{50}$ (at.%) glasses between $T_a = 400$ K and $T_b = 1$ K. The simulations do allow exploration of the effects of a range of thermal-cycling parameters.  Depending, for example, on the quench rate used to form the glass, and on the hold time at the upper temperature during cycling, the MG can undergo either rejuvenation or ageing as a result of TC.

Focusing on rejuvenation, the effects of TC are clear in progressively increased potential energy and reduced structural correlation with the as-cast glass, while the heterogeneity in the rejuvenated glass increases.  As the heterogeneity increases, the rejuvenation (quantified as the potential energy) accelerates, until levelling off at a steady-state value. Rejuvenated glasses show reduced populations of the most common atom-centred clusters, reduced elastic moduli, and an increased boson peak.  The effects in the simulations are in broad qualitative agreement with published experimental findings.

The simulations show that, during one cycle, the excitation (externally driven increase in potential energy) occurs in large measure only on heating from the lower temperature $T_b$ to the upper temperature $T_a$. We conclude that rejuvenation in TC has two causes. The first, heterogeneity in local coefficient of thermal expansion, was suggested in early work[9, 35] and computed by some of us[36], but is negligible in the present MD simulations. The second, hysteresis arising from a violation of the equipartition theorem, is found at the exceptionally high heating rates in the present simulations, but itself relies on the structural heterogeneity intrinsic to the glassy state.

The simulated Cu-Zr glasses show conclusively that cryogenic thermal cycling is capable of inducing local atomic displacements that, over progressive cycles, destroy all memory of the as-quenched structure of the glass. The properties of the TC-treated glass reach a steady state that may be relaxed or rejuvenated with respect to the as-quenched state, depending on the energy of the initial glass and the TC conditions. Simulations of the kind demonstrated in the present work, will allow exploration of the effects of TC parameters and thereby help in optimizing TC treatments.

**Methods**

**Simulation procedures and samples.** Molecular-dynamics (MD) simulations were performed with the LAMMPS package[51] for $Cu_{50}Zr_{50}$ (at.%) metallic glass with realistic embedded-atomic-method potentials[52]. The MD time step was set to 2 fs, and a constant pressure and temperature (NPT) ensemble was employed in which the temperature and pressure were controlled by the Nosé-Hoover[53, 54] and Parrinello-Rahman[55] methods respectively. To check the thermal-expansion effect in MD, we also used a constant volume and temperature (NVT) ensemble (Supplementary Fig. 8); this gave similar results. For temperature control in the heating step in TC, we also used the Berendsen thermostat (B)[56], the rescaled atomic velocity approach (R), and Langevin thermostat (L)[57]. The model system contained 10,000 atoms in a cubic box with periodic boundary conditions; we compared results with a larger system, but found no size effect. We first kept the system at 2500 K (far above the melting point) for 10 ns, and then the melt was quenched at selected rates from 2500 K to $T_a$ to prepare the initial glassy samples with different thermal histories.

**Thermal cycling (TC).** The different rates of quenching, $Q_i = 10^9, 10^{10}, 10^{11}, 10^{12}$ or $10^{13}$ K s$^{-1}$, give virtual glassy samples in different initial states. We set the upper temperature $T_a = 400$ K, which is 54% of the glass-transition temperature $T_g$ for this model system, and therefore below the temperatures normally associated with α and β thermal relaxations. Subsequent thermal cycling (Fig. 1a) consisted of cooling to the lower temperature $T_b$ (set at 1 K), holding at $T_b$, and heating back to $T_a$, at which all properties are measured. The cooling and heating were near-instantaneous, at ∼$10^{14}$ K s$^{-1}$ (chosen to be faster than the quenching rates used to form the glasses). The hold times at the lower and upper temperatures are $t_b$, set at 100 ps, and $t_a$, treated as variable to examine possible relaxation effects. The potential energy after each cycle is monitored as a function of the number of cycles $N$ and of $t_a$. TC induces changes in potential energy, $\Delta E = E_N^{t_a} - E_0$, where $E_0$ is the value for the as-quenched sample: $\Delta E > 0$ represents rejuvenation (Fig. 1a) and $\Delta E < 0$ represents relaxation (ageing). A schematic of simulation parameters in one cycle is shown in Supplementary Fig. 5.

**Mechanical properties.** To measure mechanical properties, simple shear deformation with shear rate $10^7$ s$^{-1}$ was employed to obtain the strain-stress curves (Fig. 3d) of each type of sample at 400 K. Ten independent loading runs were employed for each sample to ensure statistical validity.

Three types of samples (Supplementary Fig. 2, the three white points in the map): as-quenched ($Q_5$), aged ($t_a$ = 200 ps), and rejuvenated ($t_a$ = 10 ps) were selected to investigate the influence of TC.

**Quasi-equilibrium state ($t_b = \infty$).** As in a crystalline solid, in the glass at low temperature structural relaxation is suppressed. The thermal motion in the glass can then be regarded as atomic vibration around the inherent structure[58]. At the low temperature in our case, $T_b$ = 1 K, this atomic vibration can be approximated as a harmonic oscillator. In a classical system, the equipartition theorem[59] should work well at equilibrium, and thus for the quasi-equilibrium glassy state, we have $E_k = \Delta E_p$, where $E_k$ is the kinetic energy of the system, and $\Delta E_p$ is the difference between the potential energy at temperature $T_b$ and the potential energy of the inherent structure.

In the MD simulations, first we used a conjugate-gradient algorithm to obtain the inherent structure. We injected kinetic energy $E_k = 2k_B$ into that structure, and used the microcanonical ensemble (NVE) to relax the system. Accordingly, the system is in the quasi-equilibrium state when $E_k = \Delta E_p = 1k_B$. We have checked whether the sample starts from the quasi-equilibrium state in each cycle. In the simulation, the states obtained by TC are insensitive to the quench rate $Q_b$, holding time $t_b$ and minimization method. For the NVE relaxation, we used 2 ps in the simulation, and longer relaxation times do not change the results.

**Local property characterization.** The local parameters centred in 30×30×30 voxels (cubic, of side length ∼1 nm) were calculated by a coarse-grained method.

*Vibrational density of states (VDOS) and atomic participation ratio ($P_i$).* The dynamic matrix (DM) was calculated from energy-minimized samples, and the DM was diagonalized to obtain the vibrational density of states and the vibration mode. To extract the contribution of the Debye vibration mode, we used the reduced VDOS, $VDOS/\omega^2$. The boson peak in the low-frequency regime is shown in Fig. 4a. The participation ratio (PR) was calculated following the method used in Ref. [43], using a low 1% vibration mode. The PR of atom $i$ is given by $P_i = \sum_{N(\omega \in 1\%)} \vec{e}^i_\omega \cdot \vec{e}^i_\omega$, where $\vec{e}^i_\omega$ is the vector polarization at atom $i$ at vibration frequency $\omega$. The local PR of voxel $i$ ($\bar{P}_i$) is given by $\bar{P}_i = \frac{1}{N_i} \sum_{j \in N_i} P_j$, where $N_i$ is the number of atoms in voxel $i$, and $P_j$ is the PR of atom $j$. In glasses, the low-frequency vibration modes are localized, and the related atomic participation ratio $P_i$ can be regarded as an indicator of local structural instability. Thus, the evolution of $P_i$ could provide deep understanding about the local structural responses during rejuvenation.

***Local activated rate***. We define the local activated rate $R_i$ of voxel $i$ as

$$R_i = e^{-\frac{\Delta F_i}{k_B T_i}}$$

where $\Delta F_i$ is the local energy barrier, and $T_i$ is the local temperature, of voxel $i$.

In the 'shoving model' for molecular rearrangements in liquids[45], the relaxation energy barrier is dominated by elastic energy, and the relaxation between energy barrier and elastic energy is even true at local scale[60]. Following the formula by Wang et al.[61], here we assume that the formula still works at the local scale. Then the local energy barrier $\Delta F_i$ can be expressed as:

$$\Delta F_i = \frac{1}{11}(10 G_\infty^i + B_\infty^i)\frac{V_i}{N_i}$$

where $G_\infty^i$ and $B_\infty^i$ are the local instantaneous shear and bulk moduli, respectively; $V_i$ is the coarse-grained volume of, and $N_i$ is the number of atoms contained in, voxel $i$. Following the method used in Ref. [62], we use the local Cauchy-Born modulus to approximate the instantaneous modulus. The local temperature $T_i$ is defined by:

$$k_B T_i \equiv \frac{2}{3}E_k^i = \frac{1}{3N_i}\sum_{j\in N_i} m_j \vec{v}_j \cdot \vec{v}_j$$

where $E_k^i$ is the local kinetic energy, $\vec{v}_j$ is the velocity of atom $j$ and $m_j$ is the mass of atom $j$.

***Local non-affinity.*** Following the work of Falk & Langer[63], the non-affinity of atom $j$ can be defined as

$$D_j^2(t,\Delta t) = \frac{1}{N_j}\sum_{k\in N_j}[\vec{r}_{ij}(t+\Delta t) - \gamma_j \vec{r}_{ij}(t)]^2$$

where $\gamma_j = Y_j^{-1}\cdot X_j$, $Y_j = \sum_{k\in N_j}[\vec{r}_{jk}(t)]^T[\vec{r}_{jk}(t)]$, $X_j = \sum_{k\in N_j}[\vec{r}_{jk}(t+\Delta t)]^T[\vec{r}_{jk}(t)]$, and $\vec{r}_{jk} = \vec{r}_k - \vec{r}_j$ is the position vector of atom $j$ at time $t$. We take $D_j$, the square root of $D_j^2$, as the non-affinity of atom $j$, and local non-affinity of voxel $i$ is given by

$$\bar{D}_i(t,\Delta t) = \frac{1}{N_i}\sum_{j\in N_i} D_j(t,\Delta t)$$

We take the beginning of the holding time at $T_a$ as the reference time ($t_a = 0$), as shown by the circled region in Fig. 7a, and investigate the evolution of non-affinity with $t_a$.

***Pearson correlation coefficient.*** This coefficient is employed to characterize the correlation between $X$ and $Y$ as

$$p_{X,Y} = \frac{E[XY] - E[X]E[Y]}{\sqrt{E[X^2] - E[X]^2}\sqrt{E[Y^2] - E[Y]^2}}$$

where $E[-]$ is the expectation.

## Acknowledgements

This work was supported by The Science Challenge Project (TZ2018004), Guangdong Major Project of Basic and Applied Basic Research, China (Grant No. 2019B030302010), the NSF of China (Grant Nos. 51571111, 51271195, 51601009, U1930402) and MOST 973 Program (2015CB856800). T.C. Wang, J.-L. Barrat are thanked for discussions. A.L.G. acknowledges support from the European Research Council under the European Union's Horizon 2020 research and innovation programme (grant ERC-2015-AdG-695487: ExtendGlass). B.S.S. and P.F.G. acknowledge the computational support from the Beijing Computational Science Research Center (CSRC).

## Author contributions

P.F.G. and B.S.S conceived and designed the work. B.S.S. and P.F.G. conducted the simulations. All the authors contributed to the analysis and interpretation of the data. B.S.S., W.H.W., A.L.G. and P.F.G. wrote the manuscript.

**Competing interests:** The authors declare no competing financial interests.

**Figure Captions**

**Fig. 1** Thermal cycling (TC) and its effect on simulated $Cu_{50}Zr_{50}$ glass. **(a)** The initial glassy states are prepared by quenching at various rates $Q_i$, and then subjected to TC between $T_a$ and $T_b$. The holding times at $T_b$ and $T_a$ are $t_b$ and $t_a$ respectively. After $N$ thermal cycles, the potential energy has risen by $\Delta E$ relative to its value $E_0$ in the as-quenched glass. **(b)** The initial potential energies of the samples are indicated by the arrows. The data points show the evolution of the potential energy, $E_N^{t_a}$, as a function of cycle number $N$ for samples for which $t_a = 10$ ps or 200 ps. The initial quench rates are: $Q_1 = 10^9$ K s$^{-1}$, $Q_2 = 10^{10}$ K s$^{-1}$, $Q_3 = 10^{11}$ K s$^{-1}$, $Q_4 = 10^{12}$ K s$^{-1}$, $Q_5 = 10^{13}$ K s$^{-1}$. The other TC parameters are: $T_a = 400$ K ($= 0.54\ T_g$, where $T_g \approx 740$ K), $t_b = 100$ ps, $T_b = 1$ K, $Q_a$ and $Q_b = 2.5 \times 10^{14}$ K s$^{-1}$.

**Fig. 2** The steady-state potential energy $E_s$ of $Cu_{50}Zr_{50}$ glasses reached when they are subjected to TC. The curve of $E_s$ as a function of $t_a$ is independent of the initial (as-quenched) state, and it intersects the values (dashed lines) of the initial potential energy of glasses formed by quenching at the indicated rates $Q_i$. The downward arrows show the crossover times $t_c$. When $t_a < t_c$ the as-quenched sample is rejuvenated through TC, and when $t_a > t_c$ it is relaxed.

**Fig. 3** Effects of ageing and rejuvenation on atomic structures and mechanical properties. A $Cu_{50}Zr_{50}$ glass formed by quenching at $10^{13}$ K s$^{-1}$ is treated by TC with $t_a = 10$ ps and 200 ps to attain steady-state potential energies respectively lower than (aged) and higher than (rejuvenated) the initial as-quenched glass (Supplementary Fig. 2). **(a)** The distributions of local shear and bulk modulus for each glassy state. **(b)** The difference radial distribution functions and (in the inset) the radial distribution functions for the same three glasses. **(c)** The relative populations of the most favoured Voronoi clusters. **(d)** The strain-stress curves for deformation in simple shear at $10^7$ s$^{-1}$.

**Fig. 4** Vibrational behaviour in glasses subjected to TC. **(a)** The vibrational density of states $g(E)/E^2$ after 0, 5, 20 and 50 cycles, showing evolution (arrowed) of the boson peak. The change in potential energy $\Delta E$ (inset) increases until a steady state is achieved; the data points for 5, 20 and 50 cycles are highlighted in blue, red and green, respectively. **(b)** The Pearson correlation coefficient $p_{0,N}$ of the atomic participation ratio $P_i$ for the as-quenched state and after $N$ cycles decays with $N$. The Pearson correlation coefficient $p_{N-1,N}$ of $P_i$ for two adjacent cycled states (inset) also decays

with $N$. Three-dimensional spatial distributions of $P_i$ in the as-quenched state **(c)**, and after one cycle **(d)**; regions with notable differences between the two states are circled.

**Fig. 5** The influence of hold time $t_b$ at low temperature. **(a)** The probability distribution of local temperature $T_{b,loc}$ at low temperature $T_b = 1$ K for $t_b = 20$ ps, 200 ps, ∞ (quasi-equilibrium). **(b–d)** The corresponding 3D spatial distributions of $T_{b,loc}$ at the end of the low-temperature hold.

**Fig. 6** The progress of rejuvenation in one thermal cycle. **(a)** The evolution of potential energy in one cycle between high and low temperature, i.e. cooling from $T_a$ down to $T_b$, holding at $T_b$, then heating back to $T_a$. The evolution is shown for two values of cooling/heating rate: $10^{10}$ K s$^{-1}$ (black line) and $10^{14}$ K s$^{-1}$ (red line). **(b)** The relation between the change in potential energy $\Delta E_P$ (= $E - E_0$) and the change in kinetic energy $\Delta E_K$ during the heating stage. The solid line is followed for Cu$_{50}$Zr$_{50}$ glass and crystal CuZr at $Q_a = 10^{14}$ K s$^{-1}$ applying the Nosé-Hoover (NH) thermostat; the dashed line is followed for the MG at $Q_a = 10^{10}$ K s$^{-1}$ applying the NH thermostat, and at $Q_a = 10^{14}$ K s$^{-1}$ applying the Langevin thermostat. **(c)** The change in energy as a function of the number of cycles for the three cases in (b) with $Q_a = 10^{14}$ K s$^{-1}$.

**Fig. 7** The local activated rate ($R_i$). **(a)** The temperature evolution during the heating stage in one thermal cycle for the MG applying the Nosé-Hoover (NH) or Langevin (L) thermostats, and for CuZr crystal applying the NH thermostat. The circled region indicates the point at which the distributions of $R_i$ are calculated. Contour plots of the 3D spatial distributions of $R_i$: **(b)** for crystal CuZr, applying the NH thermostat; **(c)** for the MG, applying the L thermostat; and **(d)** for the MG, applying the NH thermostat. The simulation is controlled by the NVT ensemble and $Q_a = 2.5 \times 10^{14}$ K s$^{-1}$.

**Fig. 8.** Local activated rate ($R_i$) and local non-affinity ($\bar{D}_i$). **(a)** Distributions of $R_i$ for the MG and crystal CuZr applying the Nosé-Hoover (NH) thermostat, and for the MG applying the Langevin thermostat. The distributions are sampled at the reference point (circled) in Fig. 7a, over a time interval of 1 ps. **(b)** The correlation between $\bar{D}_i$ and $R_i$ for the three cases in (a): NH MG (red), Langevin MG (orange), and NH crystal (green), with Pearson correlation coefficients of 0.6, 0.06, and –0.03 respectively.

**Fig. 9** Evolution of the local non-affinity ($\bar{D}_i$) during the hold at $T_a$. Contour plots of the 3D spatial

distributions of local non-affinity between the start of the hold (circled region in Fig. 7a) and after hold times $\Delta t$ of **(a)** 0.2 ps, **(b)** 1 ps, **(c)** 4 ps, and **(d)** 80 ps.

**Figure 1**

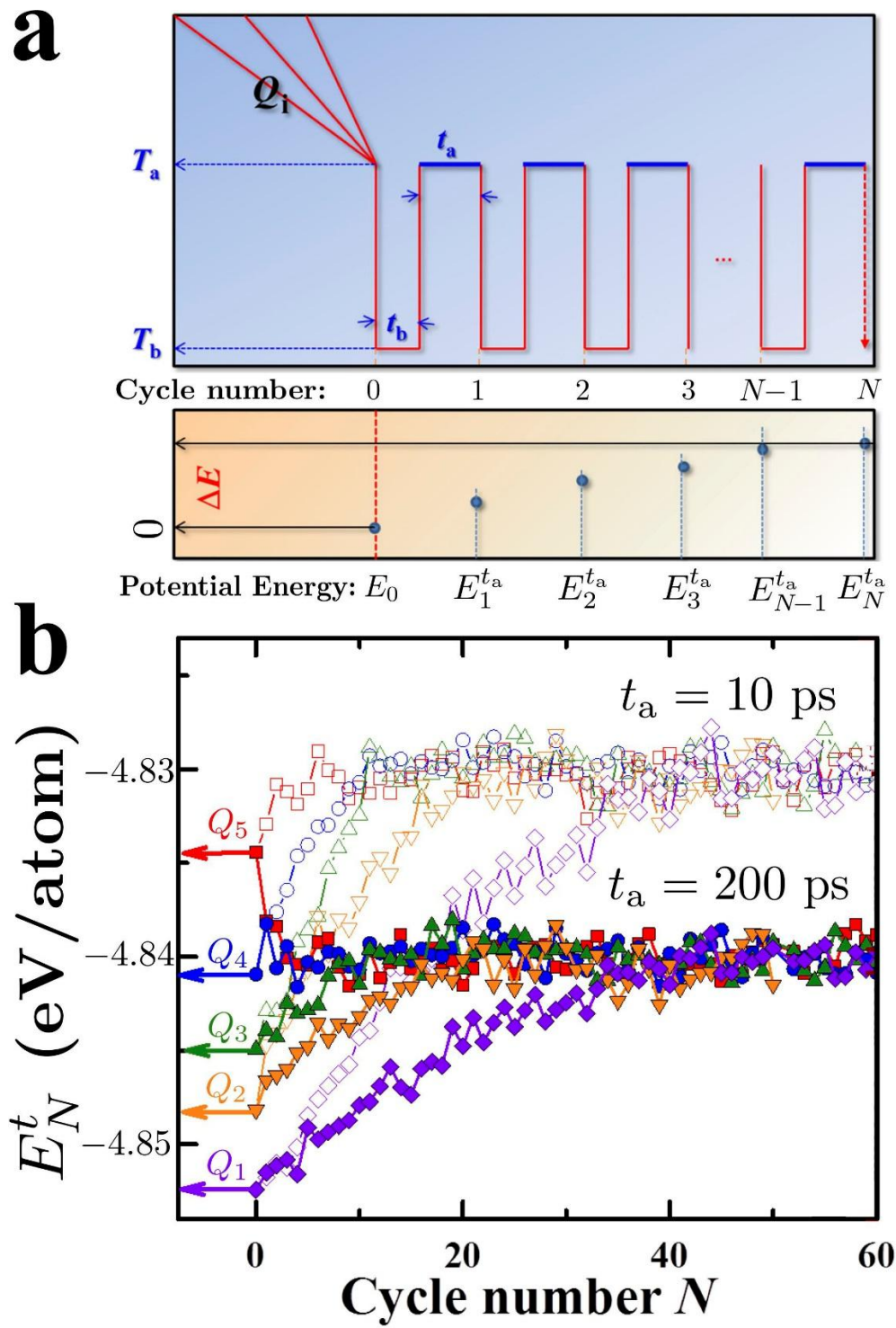

Figure 1. Shang et al

**Figure 2**

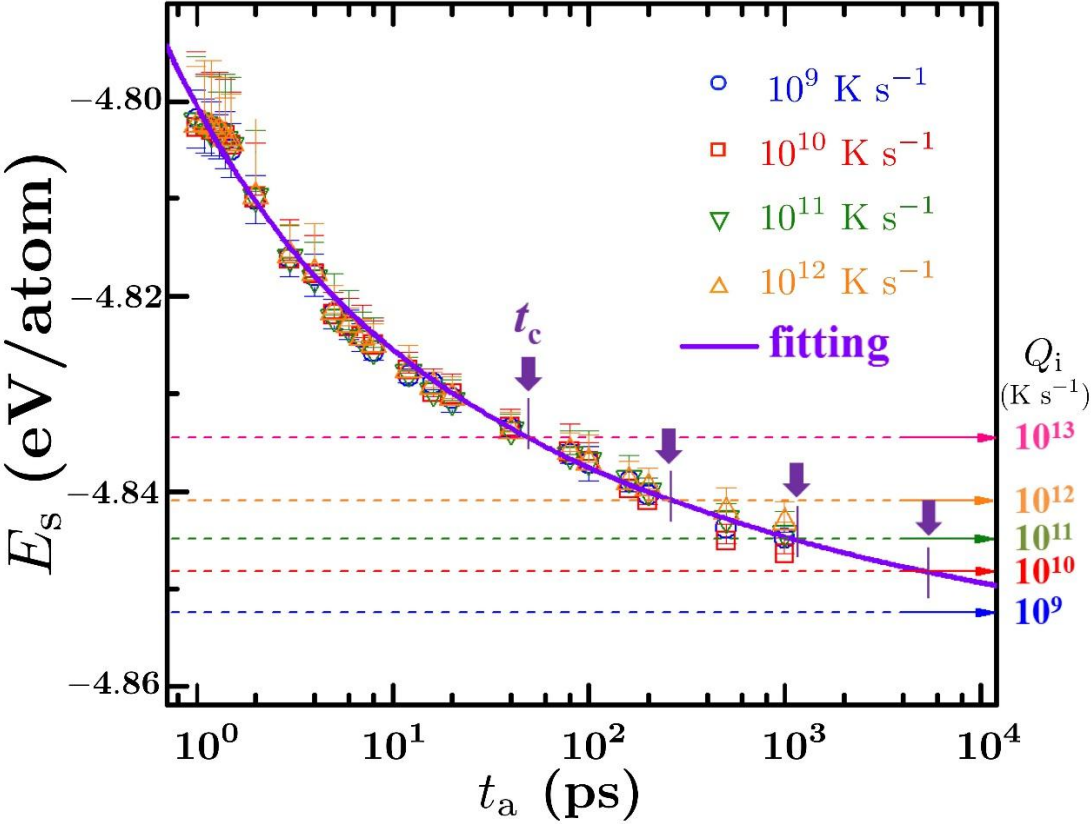

**Figure 2，Shang et al**

**Figure 3**

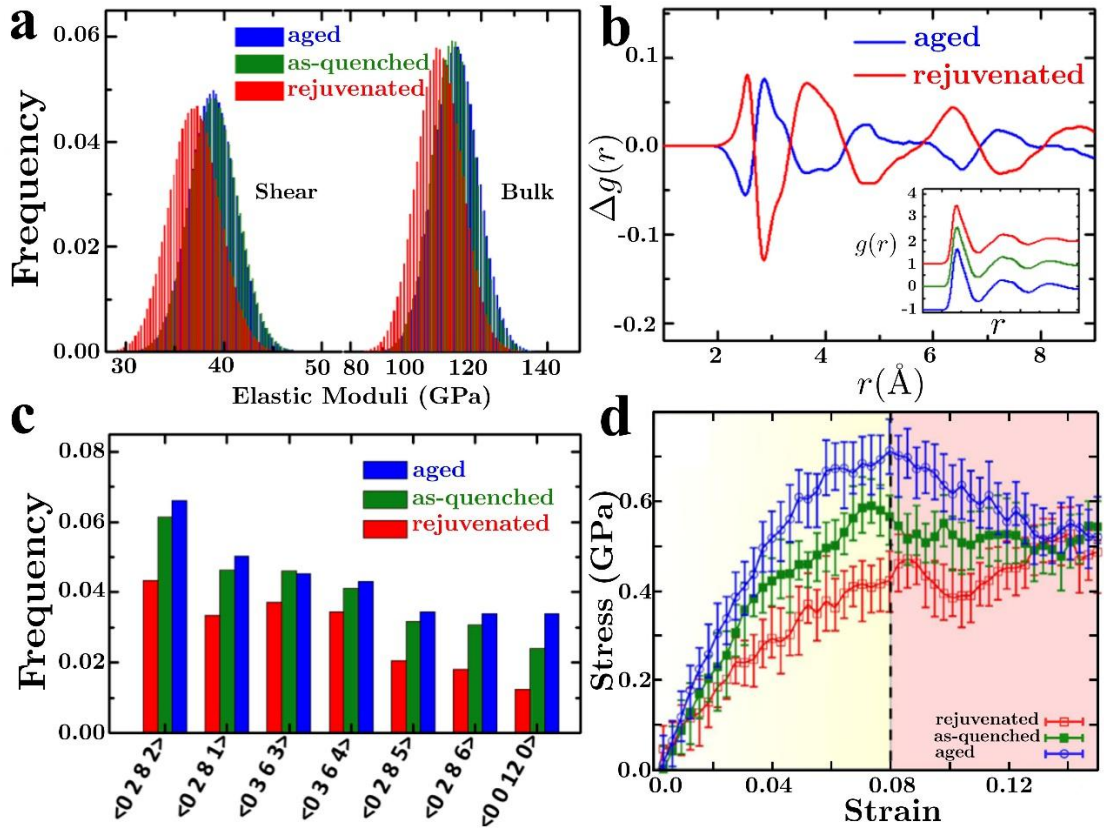

Fig. 3 Shang et al

Figure 4

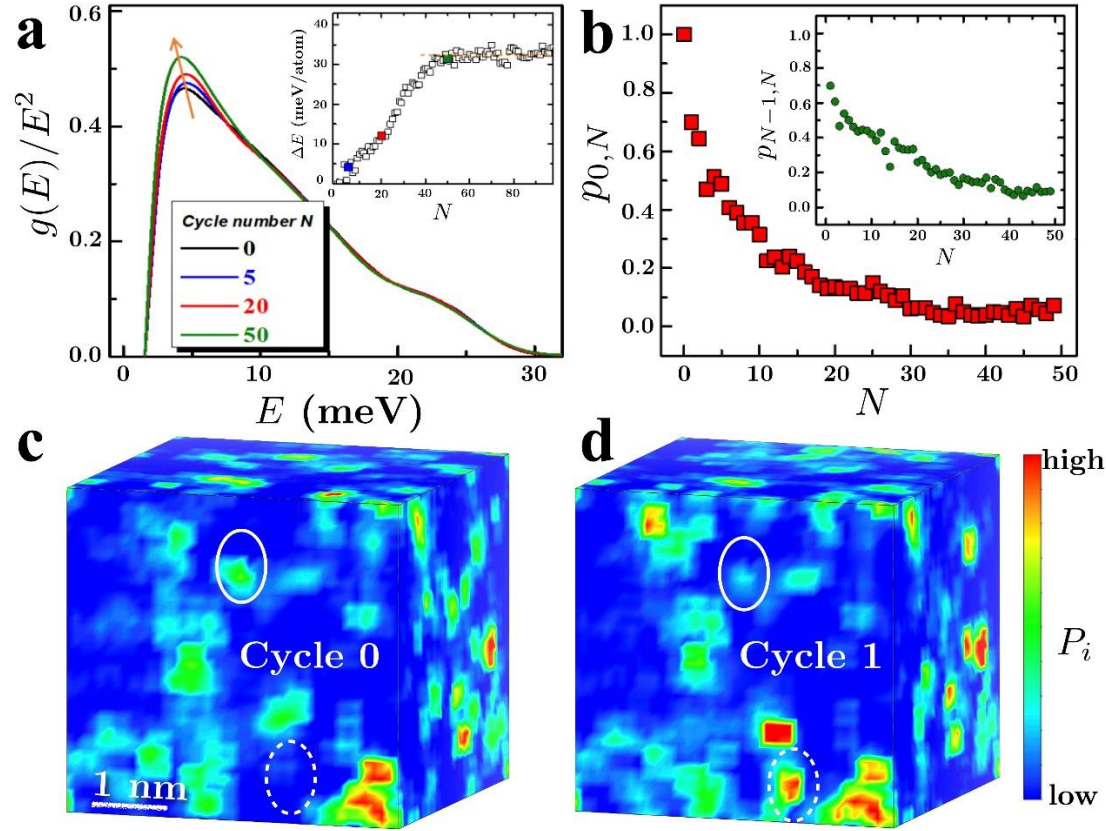

Fig. 4 Shang et al

**Figure 5**

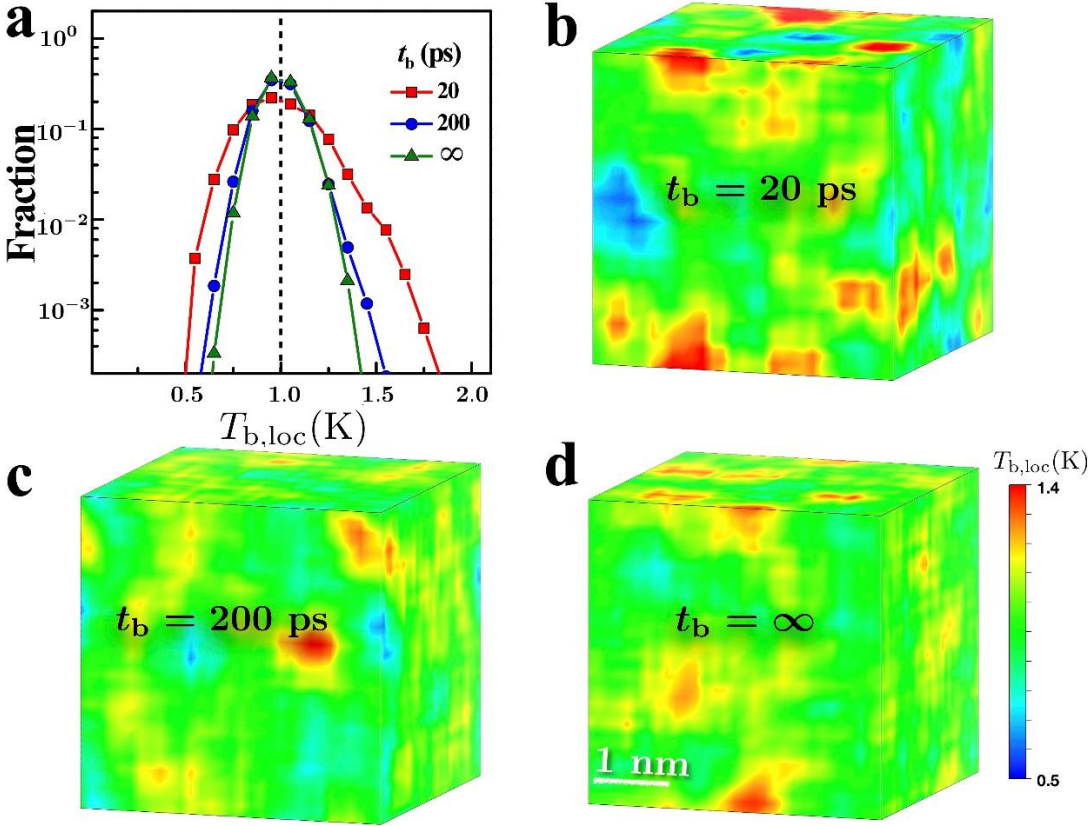

Fig. 5 Shang et al

Figure 6

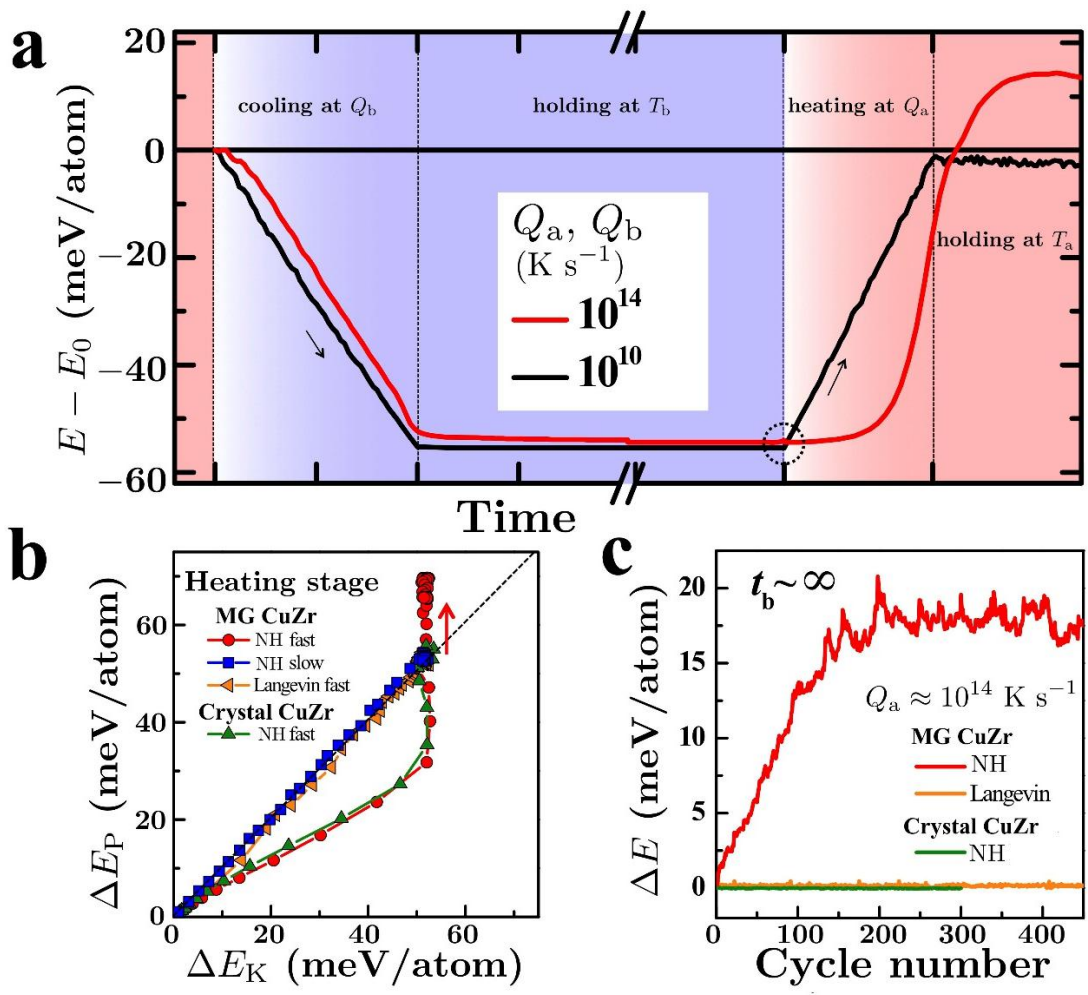

Fig. 6 Shang et al

**Figure 7**

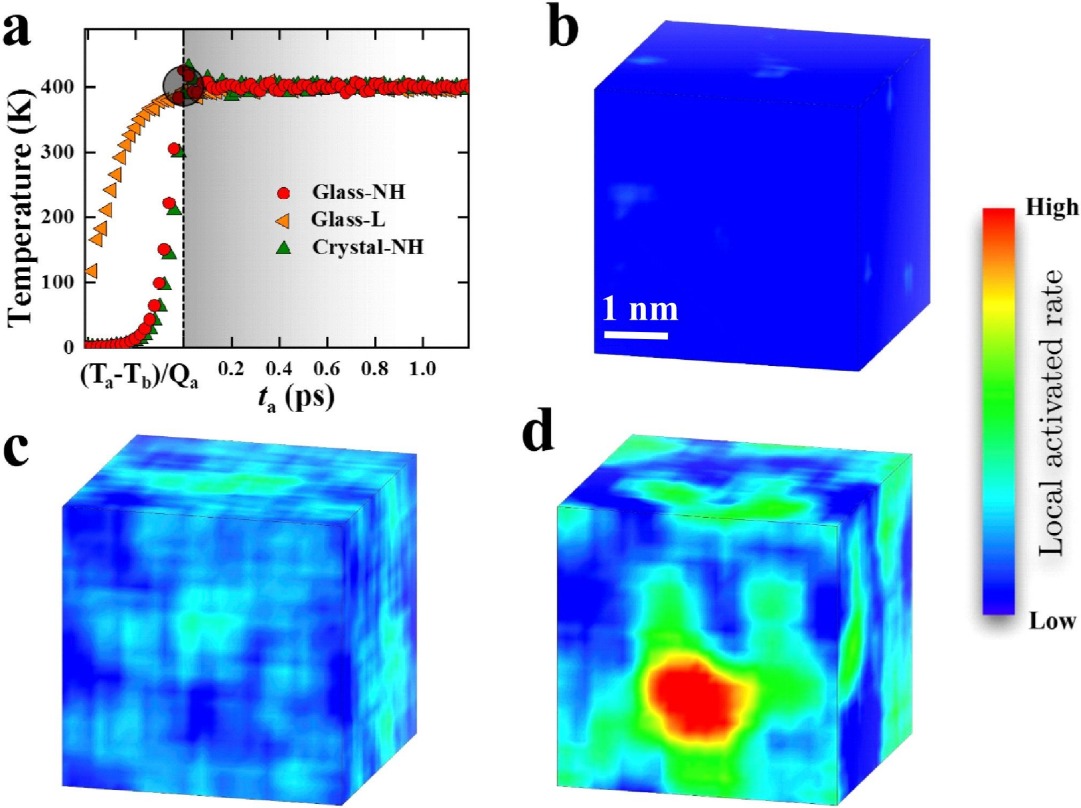

**Fig. 7 Shang et al**

Figure 8

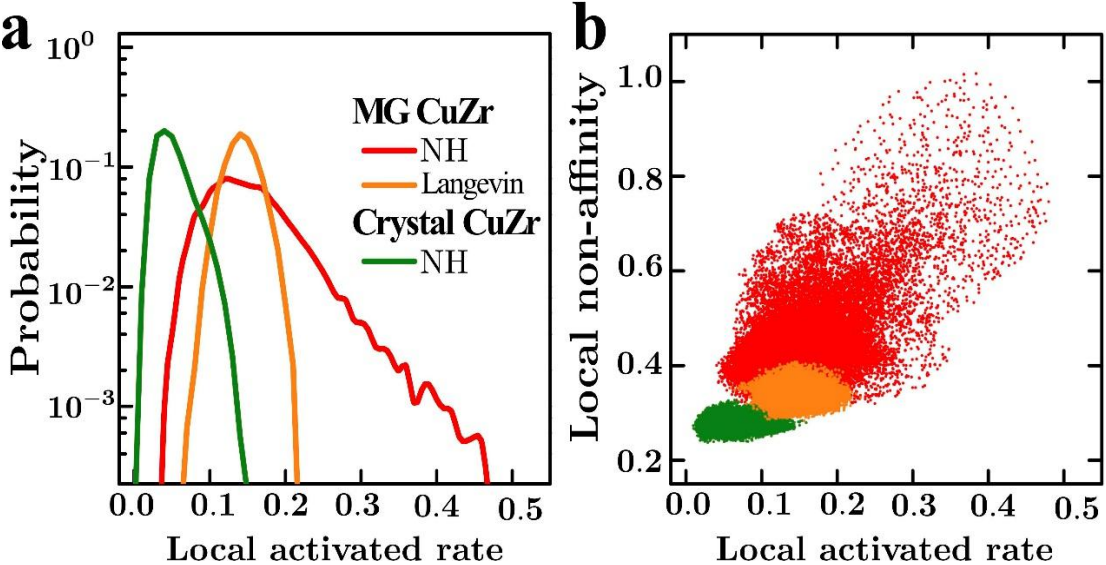

Fig.8 Shang et al

**Figure 9**

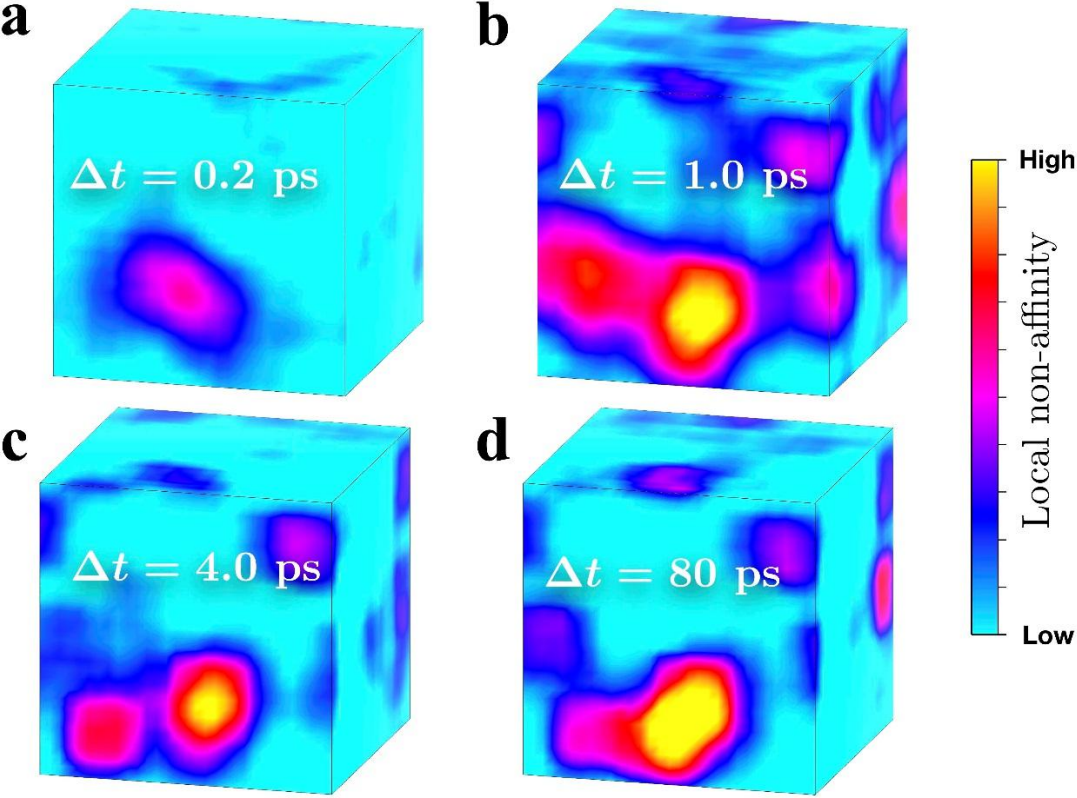

**Fig.9 Shang et al**

# SUPPLEMENTARY MATERIALS:

# Atomistic Modelling of Thermal-Cycling Rejuvenation in Metallic Glasses

Baoshuang Shang,[1,2] Weihua Wang,[3,2] Alan Lindsay Greer,[4] and Pengfei Guan[1*]

[1] *Beijing Computational Science Research Center, Beijing 100193, China*

[2]*Songshan Lake Materials Laboratory, Dongguan 523808, China*

[3]*Institute of Physics, Chinese Academy of Sciences, Beijing 100190, China*

[4]*Department of Materials Science and Metallurgy, University of Cambridge,
27 Charles Babbage Road, Cambridge CB3 0FS, UK.*

*pguan@csrc.ac.cn

1. **Ageing versus rejuvenation during thermal cycling**

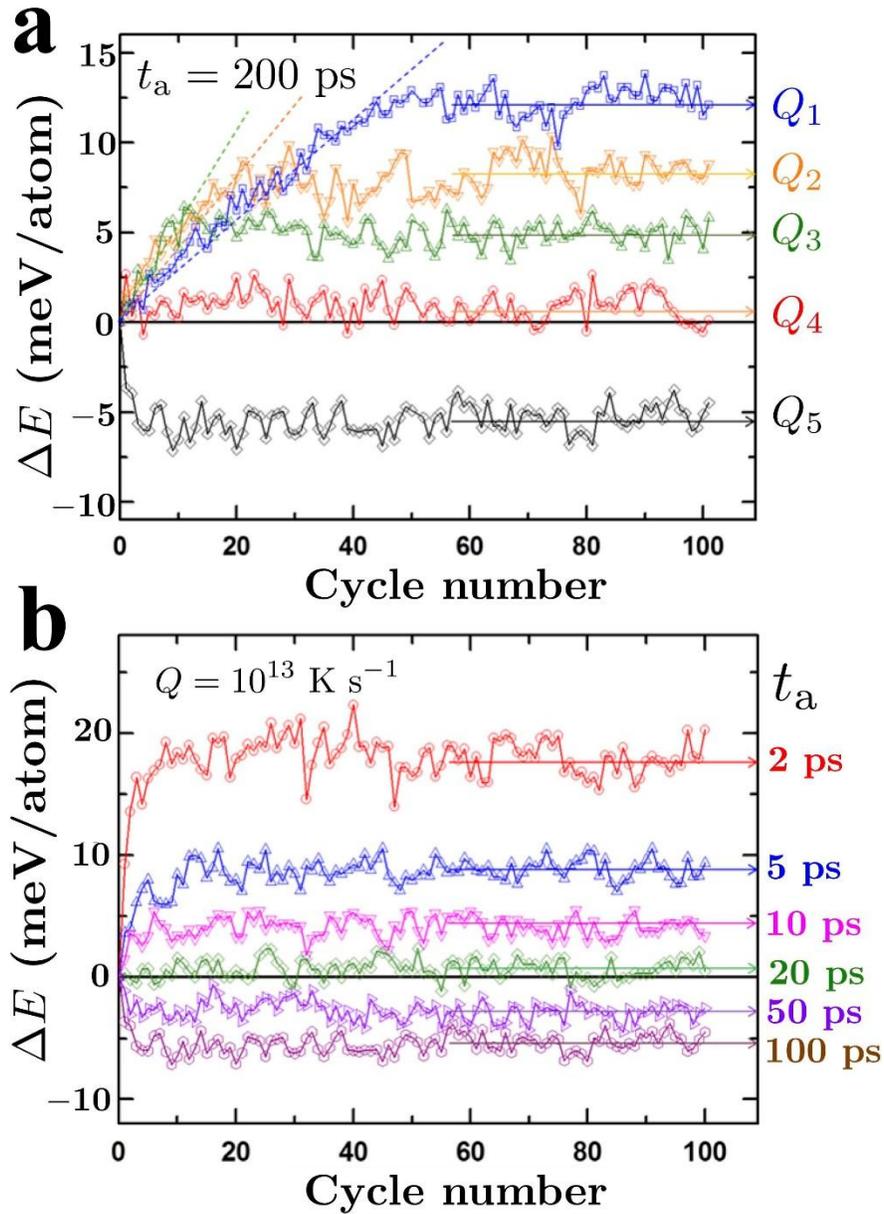

**Fig. S1** The change in potential energy of $Cu_{50}Zr_{50}$ MG as a function of TC. **(a)** The change in energy $\Delta E$ as a function of the number of cycles $N$ with $t_a$ = 200 ps for samples prepared by quenching at different rates from $Q_1 = 10^9$ K s$^{-1}$ to $Q_5 = 10^{13}$ K s$^{-1}$. **(b)** $\Delta E$ as a function of $N$ for samples prepared at $Q_5$ and held for different times $t_a$ at the upper temperature. Reheating/cooling rate $Q_a = Q_b = 2.5 \times 10^{14}$ K s$^{-1}$, $T_b$ = 1 K, $t_b$ = 200 ps and $T_a$ = 400 K.

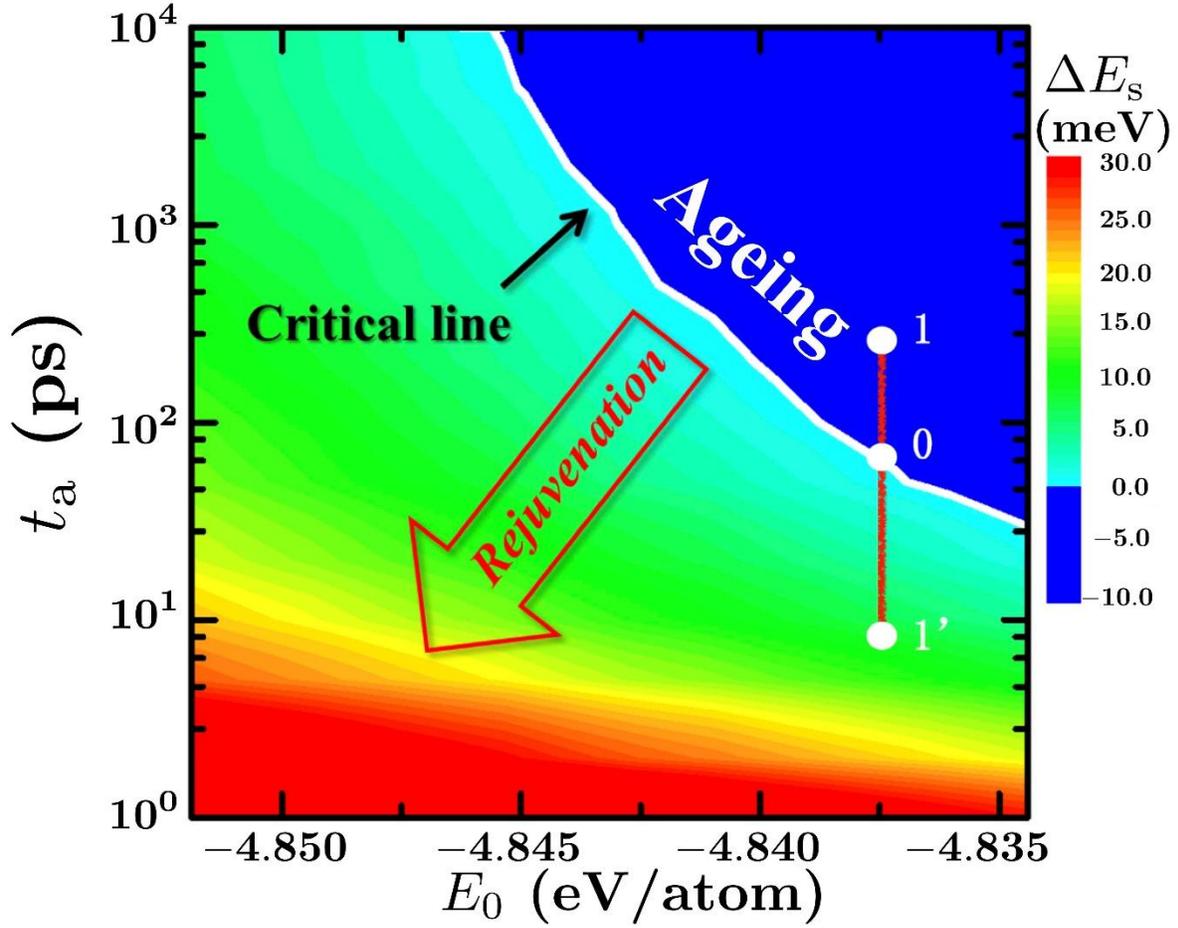

**Fig. S2** Map of the potential energy change $\Delta E$ induced by TC. The value of $\Delta E$ depends on the energy $E_0$ of the initial as-quenched glass and on the hold time $t_a$ at the upper temperature. The three points on the map represent: (1) a relaxed (aged) sample for which $\Delta E$ is negative, (0) the as-quenched sample, and (1') a rejuvenated sample for which $\Delta E$ is positive. Reheating/cooling rate $Q_a = Q_b = 2.5 \times 10^{14}$ K s$^{-1}$, $T_b = 1$ K, $t_b = 200$ ps and $T_a = 400$ K.

## 2. VDOS of thermally cycled samples

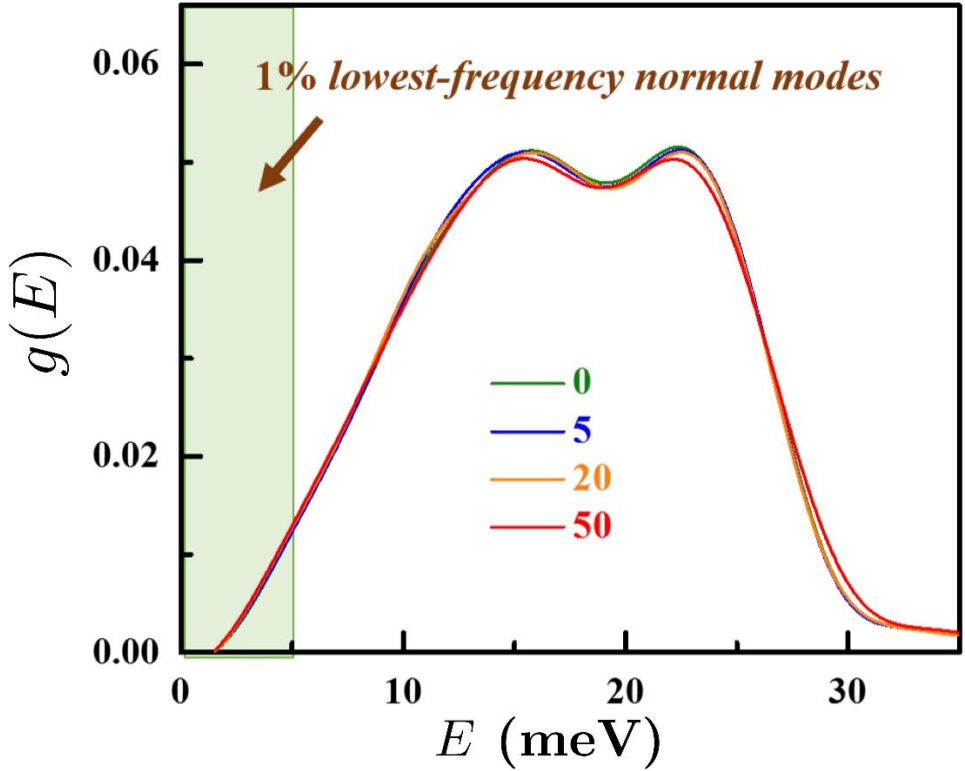

**Fig. S3** The distribution of the vibrational density of states (VDOS) for an as-quenched sample and samples after 5, 20 and 50 thermal cycles. The shaded area shows the fraction of the distribution used to calculate the atomic participation ratio $P_i$. The default values are: reheating/cooling rate $Q_a = Q_b = 2.5 \times 10^{14}$ K s$^{-1}$, $t_b = 200$ ps, $t_a = 4$ ps, $T_b = 1$ K and $T_a = 400$ K.

## 3. The evolution of soft spots in TC

In TC, the soft spots in metallic glasses can be activated, and during the atomic rearrangement they can be produced or annihilated. There are examples of both disappearance and generation of soft spots (dashed circles in Fig. S4), but overall an increase in their population. With increasing number of cycles the increased population reaches saturation. For the initial few cycles (Fig. S4), most soft spots are still stable, indicating that the rejuvenation is quite localized. In contrast, when the population density of soft spots reaches saturation, the spatial distribution is noticeably shuffled; this is caused by a sequence of interactions between soft spots.

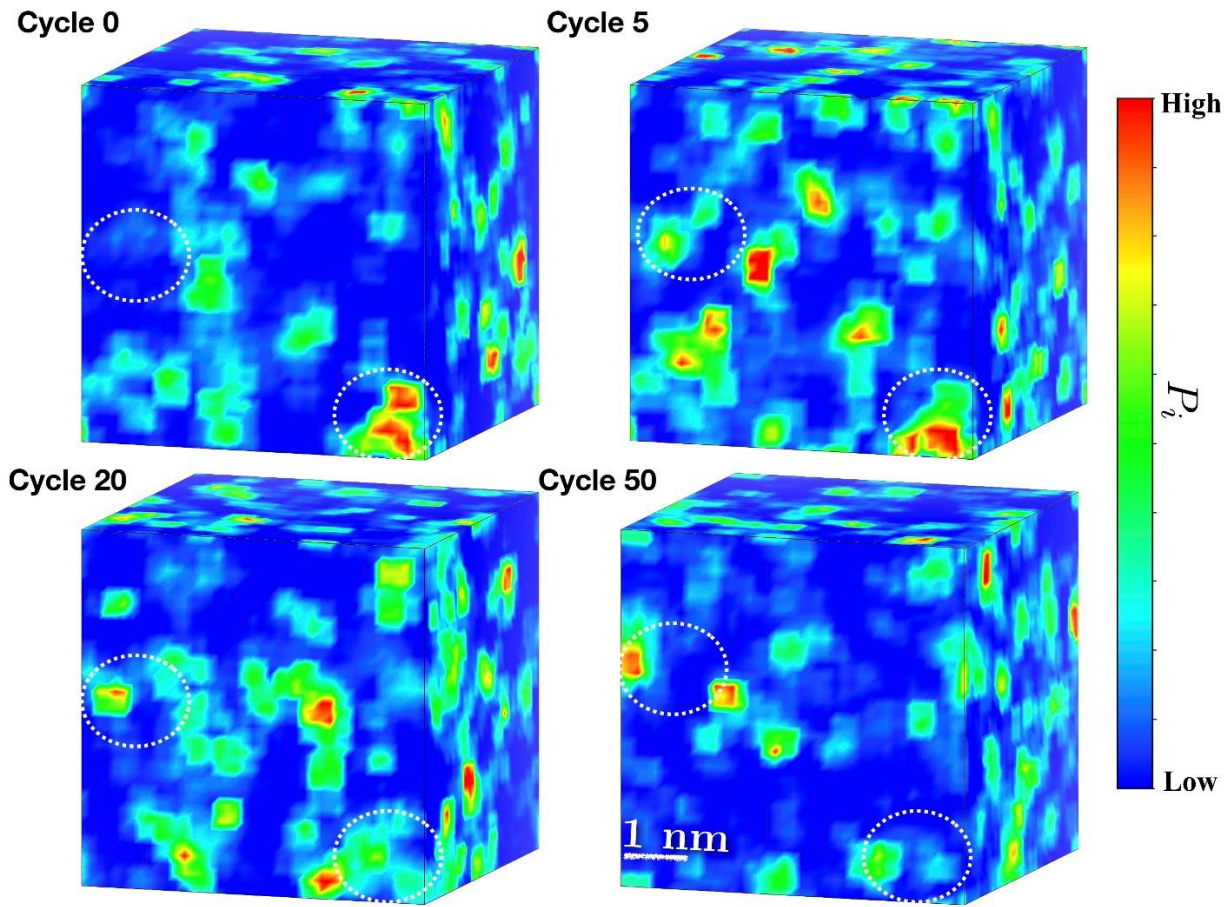

**Fig. S4** Evolution of the distribution of the atomic participation ratio ($P_i$) upon TC. Contour plots of the 3D spatial distributions of $P_i$: **(a)** in the as-quenched state, and in states after **(b)** five, **(c)** 20, and **(d)** 50 cycles. The initial glass is formed by quenching at $Q_1 = 10^9$ K s$^{-1}$. The default values are: reheating/cooling rate $Q_a = Q_b = 2.5 \times 10^{14}$ K s$^{-1}$, $t_b = 200$ ps, $t_a = 4$ ps, $T_b = 1$ K and $T_a = 400$ K.

## 4. The definition of key parameters in TC

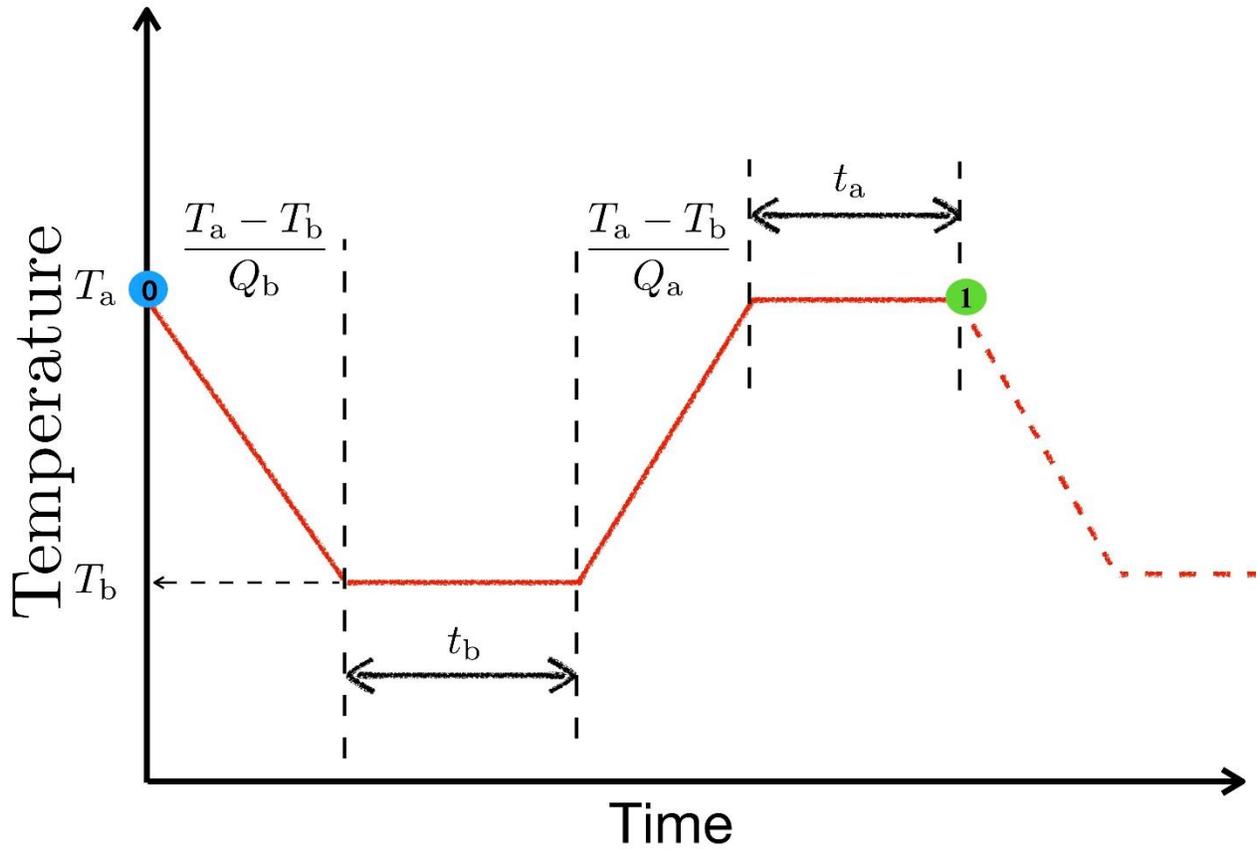

**Fig. S5** The parameters in thermal cycling (TC). Each cycle starts from the blue point (state 0) and is characterized by the cooling rate $Q_b$, the hold time $t_b$ at the lower temperature $T_b$, the heating rate $Q_a$, and the hold time $t_a$ at the upper temperature $T_a$, and ends at the green point (state 1).

## 5. The influence of key factors on the rejuvenation

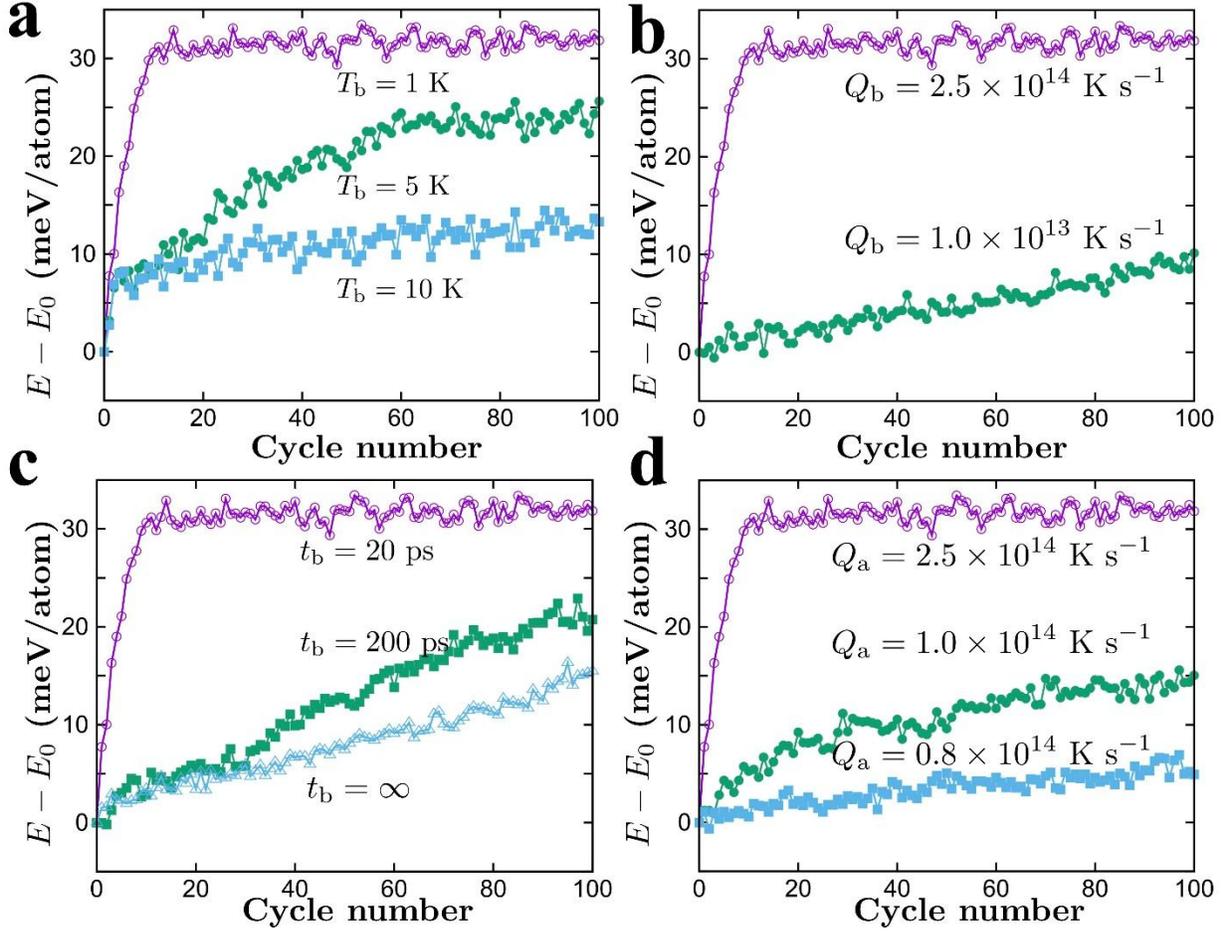

**Fig. S6** Factors influencing the extent of rejuvenation induced by TC. Key factors are: **(a)** the lower temperature $T_b$, **(b)** the cooling rate $Q_b$, **(c)** the hold time $t_b$ at the lower temperature, and **(d)** the heating rate $Q_a$.

## 6. The influence of the thermostat

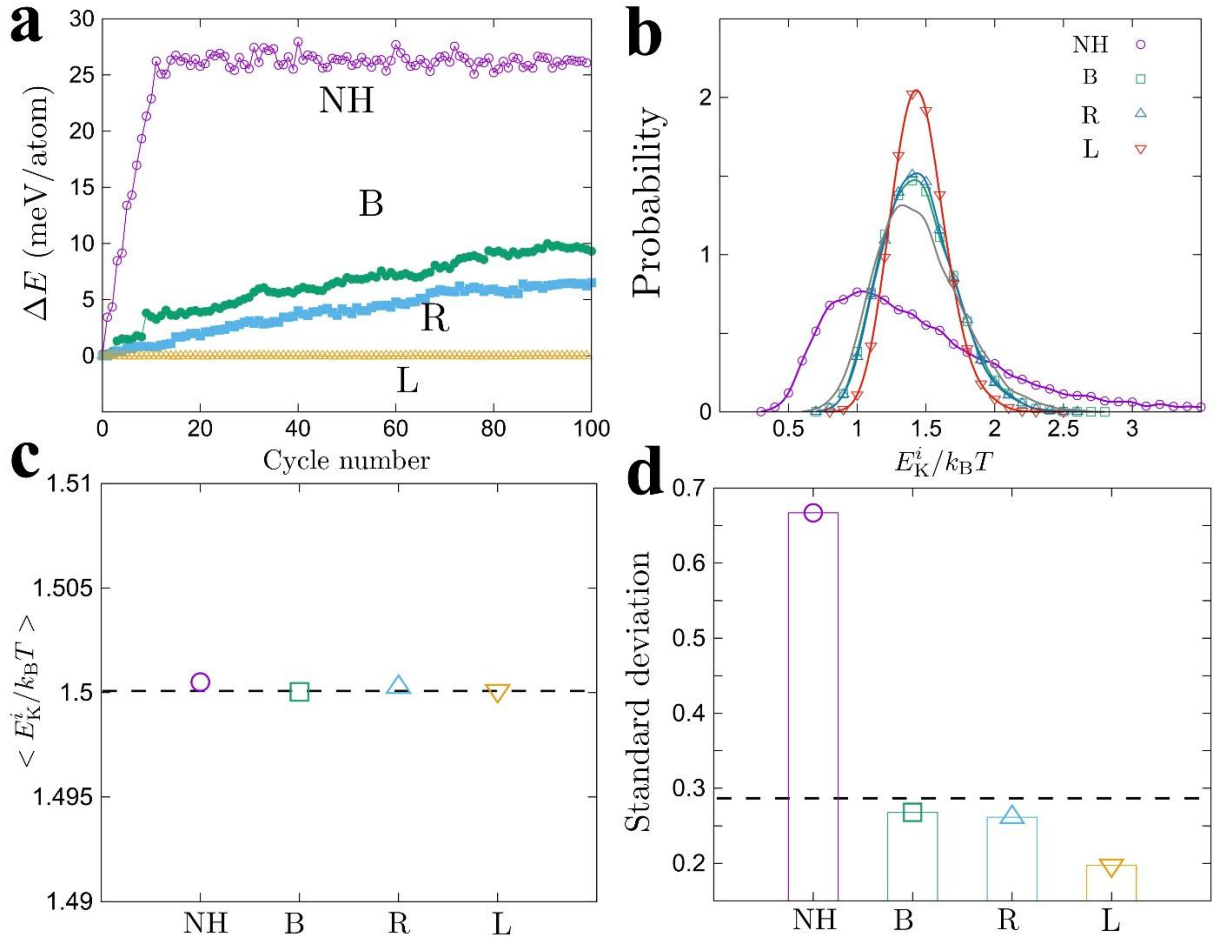

**Fig. S7** The influence of the thermostat in simulations of thermal cycling of MG. **(a)** Evolution of potential energy at 1 K with cycle number for Nosé-Hoover (NH), Berendsen (B), rescaled (R) and Langevin (L) thermostats. **(b)** The distributions of local kinetic energy $E_K^i/k_B T$ at the start of the first heating stage under different thermostats; the initial state (atomic configuration and atomic kinetic energy distribution) at the start of the first hold $t_b$ is the same for all thermostats. **(c)** The average of the local kinetic energy $\langle E_K^i/k_B T \rangle$ for different thermostats. The dashed line shows the value of $\langle E_K^i/k_B T \rangle$ in the initial state, which suggests that $E_K^i$ fluctuates around 1 K for all thermostats. **(d)** The standard deviation of the distribution of the local kinetic energy $E_K^i/k_B T$. The dashed line presents the standard deviation of the initial state, the spatial distributions of $E_K^i$ are shown in Fig. 5b. The default values are: reheating/cooling rate $Q_a = Q_b = 2.5 \times 10^{14}$ K s$^{-1}$, $t_b = 200$ ps, $t_a = 4$ ps, $T_a = 400$ K and $T_b = 1$ K.

As shown in Fig. S7a, the degree of rejuvenation is highly dependent on the choice of thermostat. In MD simulations, the thermostat is intended to mimic the processes of 'equilibration' in the physical world. It is difficult to say which of various possible thermostats best represents the physical world in a highly non-equilibrium heating process. However, the basic principle should be consistency between MD and the physical world with regard to different degrees of rejuvenation.

When the distribution of local kinetic energy at the start of the *first heating stage* is more heterogeneous (Fig. S7b, d), the degree of rejuvenation during heating is greater. For the Nosé-Hoover thermostat, the heterogeneity is greatly enlarged after heating while, in contrast, the Berendsen and the rescaled thermostats maintain the initial heterogeneity. Since the Langevin thermostat controls the temperature by a stochastic equation, the heterogeneity of local kinetic energy is maintained at the lowest level, and there is no correlation with the initial state. The mean value of the local temperature is identical in all cases. Thus, there is a special role for the heterogeneity of local kinetic energy, which, in addition to the glassy structure itself, can also drive the rejuvenation or relaxation.

## 7. The effect of bulk thermal expansion in simulations

Due to the heterogeneous structure of metallic glasses, non-uniform thermal expansion has been considered as the origin of rejuvenation in cryogenic thermal cycling (CTC). In our MD simulations and previous work [11], we find that the probability of activation by the local thermal expansion mismatch is quite small. If we suppress bulk thermal expansion during TC by fixing the sample volume, the efficiency of rejuvenation does not change significantly (Fig. S8), that is to say, bulk thermal expansion is unnecessary for CT induced rejuvenation. There is a role for structural heterogeneity, but not for overall volume change.

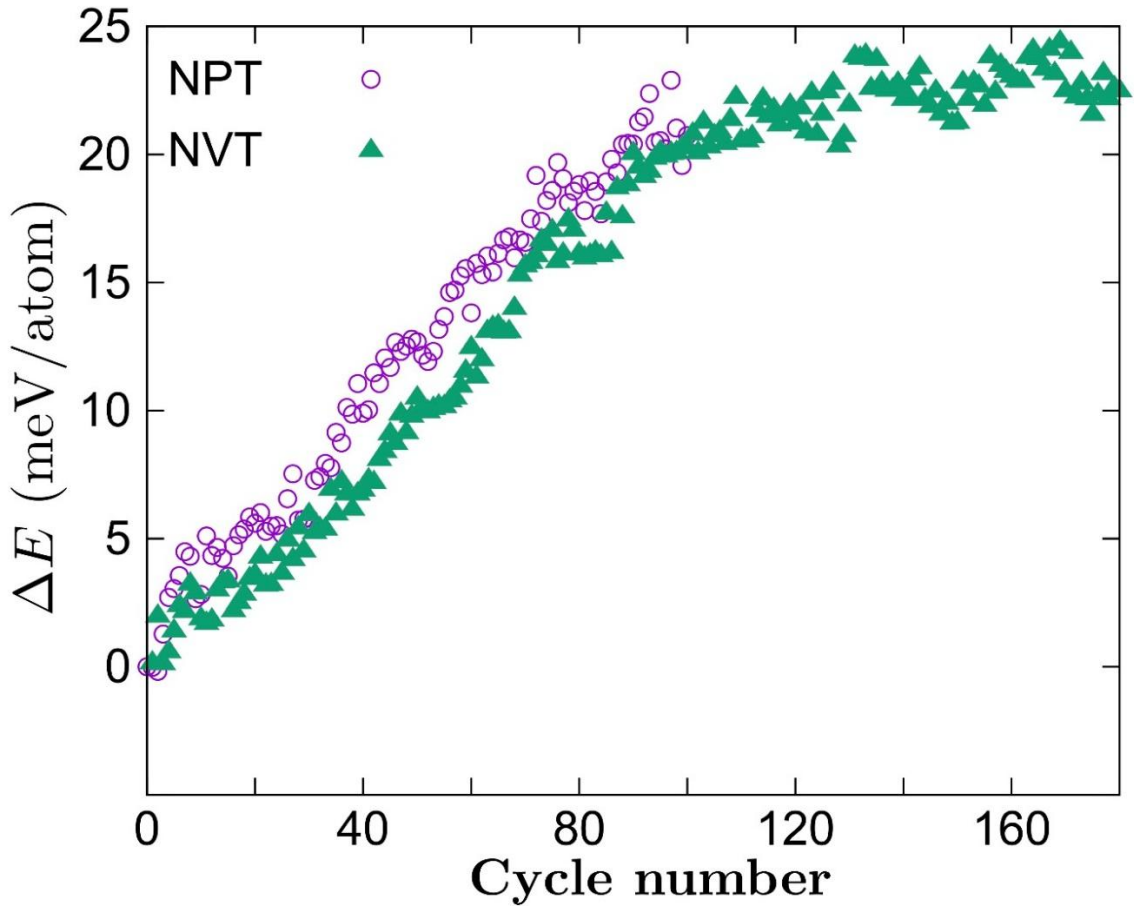

**Fig. S8** Comparison of the NVT and NPT ensembles in simulations of the effects of TC. The increase in potential energy $\Delta E$ is similar in each case. The temperature is controlled by the Nosé-Hoover thermostat, the pressure is controlled by the Parrinello-Rahman barostat, $T_a = 400$ K, $T_b = 1$ K, $Q_a = Q_b = 2.5 \times 10^{14}$ K s$^{-1}$, $t_b = 200$ ps and $t_a = 4$ ps.